\documentclass[12pt]{iopart}
\usepackage[latin1]{inputenc}
\usepackage{psfrag,bm,graphicx}
\newcommand{\average}[1]{{\left\langle{#1}\right\rangle}}
\renewcommand{\vec}[1]{{\bm{#1}}}
\newcommand{\D}{\rmd}
\newcommand{\E}{\rme}

\begin{document}
% Journal identifier can be put here if required, e.g.
%\jl{14}

\title[Equipartition in gas mixtures]{On the 
equipartition of kinetic energy in an ideal gas mixture}

\author{L. Peliti}

\address{Dipartimento di Scienze Fisiche, INFN--Sezione di Napoli,
CNISM--Unità di Napoli\\
Università ``Federico~II'', Complesso Monte S. Angelo, 80126--Napoli (Italy)}
\ead{peliti@na.infn.it}

\begin{abstract}
A refinement of an argument due to Maxwell for the equipartition 
of kinetic energy in a mixture of ideal gases with different masses
is proposed. The argument is elementary, yet it may work as an illustration of
the role of symmetry and independence postulates in kinetic theory.
\end{abstract}

\pacs{05.20.Dd}

\submitto{\EJP}

\maketitle

\section{Introduction}
The intuitive appeal of the classical kinetic theory of gases
in the teaching of the basics of statistical mechanics is
undeniable. However, one often finds that it is hard to
continue along the kinetic trail without becoming entangled
either in too difficult mathematics, or in too subtle arguments,
which make the subject less glamorous than it should. 
I found the sections of the Feynman lectures dedicated
to the kinetic theory~\cite[secs.~39,~40]{Feynman} remarkable
in balance and scope, in the attempt to derive the essentials
of the Gibbs approach to statistical mechanics by kinetic
considerations which are usually associated with Boltzmann.
One of the original points of Feynman's approach is the fact that 
equipartition is taken as a starting point to derive
Maxwell's velocity distribution and the Boltzmann factor,
rather than as a consequence. Equipartition could be inferred
by putting together the expression of the gas pressure
as a function of the mean kinetic energy---obtained by the
classic Bernouilli~\cite[sec.~X]{Bernouilli} reasoning---and 
the ideal gas law. However
a \textit{derivation} of equipartition requires something
more. In particular it is necessary to show that
if a mixture of gases of different masses is held
in the same vessel, the mean kinetic energy per particle
for each kind of gas is the same at equilibrium.
The argument given by Feynman is not totally convincing, 
as he himself remarks:
\begin{quote}
This argument, which was the one used by Maxwell, involves some
subtleties. Although the conclusions are correct, the result does
\textit{not} purely from the considerations of symmetry that we used
before, since, by going to a reference frame moving through the gas,
we may find a distorted velocity distribution. We have not found a
simple proof of this result.   
\end{quote}
It is my aim in the present note to provide such a simple proof---at the
level of rigor of the remainder of Feynman's discussion. My argument
is basically a refinement of the admittedly weak one put 
forward by Maxwell in his 1860 memoir~\cite{Maxwell60}, 
rather than the more sophisticated one
introduced in the 1867 memoir~\textit{On the Dynamical Theory of 
Gases}~\cite{Maxwell67}.
Both papers are accessible in S. G. Brush's collection 
\textsl{Kinetic Theory}~\cite[vol.~1,~n.~10;~vol.~2,~n.~1]{Brush}.

\section{Maxwell's argument of 1867}
In his 1867 paper, Maxwell considers the velocity distribution for
particles of the two kinds at equilibrium. Let us denote by $\vec{v}_1$
the velocity of a particle of the first kind and by $\vec{v}_2$ that
of one of the second kind. After the collisions, let us denote
the respective velocities by $\vec{w}_{1,2}$. Then, at equilibrium, the
number of collisions going from $(\vec{v}_1,\vec{v}_2)$ 
to  $(\vec{w}_1,\vec{w}_2)$ should be balanced by those 
going from  $(\vec{w}_1,\vec{w}_2)$ to $(\vec{v}_1,\vec{v}_2)$.
Maxwell argues in the following way that the balance
should be \textit{detailed}, velocity pair by
velocity pair (I slightly changed the notations):
\begin{quotation}
     Suppose that the number of molecules having
velocity $\vec{v}'$ increases at the expenses of $\vec{v}$.
Then since the total number of molecules corresponding to $\vec{v}'$
remains constant, $\vec{w}$ must communicate as many to
$\vec{v}''$, and so on till they return to $\vec{v}$.

Hence if $\vec{v}$, $\vec{v}'$, $\vec{v}''$ be a series of velocities,
there will be a tendency of each molecule to assume the
velocities $\vec{v}$, $\vec{v}'$, $\vec{v}''$, etc.\ in order,
returning to $\vec{v}$. Now it is impossible to assign a 
reason why the successive velocities of a molecule should
be arranged in this cycle, rather than in the reverse order.
If, therefore, the direct exchange between $\vec{v}$ and
$\vec{v}'$ is not equal, the equality cannot be preserved by exchange
in a cycle.
\end{quotation}
If the velocities of particles of the two kinds are independent,
and if we denote by $f_{1,2}(\vec{v})$ the respective velocity
distributions, we have at equilibrium
\begin{equation}\label{balance:eq}
  f_1(\vec{v}_1)f_2(\vec{v}_2)=f_1(\vec{w}_1)f_2(\vec{w}_2).
\end{equation}
But the only connection between the pairs $(\vec{v}_1,\vec{v}_2)$ 
and $(\vec{w}_1,\vec{w}_2)$ is that the total kinetic energy is
conserved. Thus both sides of eq.~(\ref{balance:eq}) can only
depend on the total kinetic energy. This implies that $f_i(\vec{v})$
($i=1,2$) can only depend on the kinetic energy of each particle, and 
moreover that it must be of the form
\begin{equation}
  f_i(\vec{v})\propto \E^{-\frac{\beta}{2}m_iv^2 },
\end{equation}
where $m_i$ is the mass of particles of kind~$i$,
and $\beta$ is a positive constant. Equipartition follows immediately.

This is an argument of great elegance, but it is probably a bit too abstract
for an introductory lecture. In particular, it focuses on a necessary
condition for equilibrium, while it could be more appealing to consider
at least hypothetically the approach to equilibrium. 
Such an argument was first considered by Maxwell in his 1860 memoir.

\section{Maxwell's argument of 1860}
In his 1860 paper, Maxwell considers the effect of one collision
between one particle of kind~1 and one of kind~2, animated by
velocities equal, in modulus, to the mean velocity of each kind,
and whose directions are perpendicular to each other. Indicating
by $v_i$ ($i=1,2$) the modulus of the velocity of the particle
of kind $i$, let us denote by $w_i$ ($i=1,2$) the corresponding
moduli after impact. Then, by solving the problem of impact between hard
spheres in this geometry, Maxwell shows that
\begin{equation}
  m_1 w_1^2-m_2 w_2^2 = \left(\frac{m_1-m_2}{m_1+m_2}\right)^2
  \left(m_1 v_1^2-m_2 v_2^2\right).
\end{equation}
Thus the difference between the kinetic energies of the two kinds
of particles is reduced by such an impact, by a ratio that depends
only on the masses of the two particles. Maxwell then
argues that it should vanish at  equilibrium.

S. G. Brush~\cite[\S~10.4]{BrushKM} makes the following comment:
\begin{quote}
     It seems amazing to me that Maxwell should have thought he
was proving a tendency toward equalization of kinetic energies
by this argument, or that any of his contemporaries who
bothered to examine the argument in detail should have accepted
it. All Maxwell has done is to pick out
one very special kind of collision for which the kinetic energies become
more nearly equal and then claim that the same result will follow for 
\textit{all} collisions.
\end{quote}

It seems clear to me that one cannot hope to derive the tendency toward
equipartition without some additional statistical assumptions: after all,
microscopic reversibility stands in the way. But it \textit{is} possible
to refine this argument in order to show that if the velocity distribution
of the particles is such that the center-of-mass motion is correlated
with the relative motion of colliding particles, this correlation is
reduced for the outgoing particles after the collision. Then, if the velocities
of the colliding particles are independent (the molecular chaos hypothesis)
this result implies that the difference between the kinetic energies
of the particles of the two kinds are indeed reduced by collisions.
This is explained in the next section.

\section{Equipartition in a gas mixture}
Let us consider a mixture of two gases, kind~1 with mass $m_1$ and kind~2 
with mass $m_2$. We assume that the range of interactions among the particles
is finite, and much smaller than the interparticle distance, so that
it is safe to assume that the particles do not interact among 
themselves except for the very short time in which they collide.
The collisions are elastic, and conserve the momentum. Let us now
consider a collision between a particle of kind~1, animated by
velocity $\vec{v}_1$, and one of kind~2, animated by velocity $\vec{v}_2$.
Let us denote by $\vec{w}_1$ and $\vec{w}_2$ the respective velocities 
after the collision.

The laws of conservation of momentum and energy stipulate
\begin{eqnarray}
  m_1\vec{v}_1+m_2\vec{v}_2&=&m_1\vec{w}_1+m_2\vec{w}_2;\\
  \frac{1}{2}m_1 v_1^2+\frac{1}{2}m_2 v_2^2&=&\frac{1}{2}m_1 w_1^2
  +\frac{1}{2}m_2 w_2^2.\label{kinerg:eq}
\end{eqnarray}
As a consequence of these relations, the absolute value of the relative
velocity remains the same before and after the collision. Setting
$\vec{V}=\vec{v}_2-\vec{v}_1$ and $\vec{W}=\vec{w}_2-\vec{w}_1$, we
have
\begin{equation}\label{relvec:eq}
  |\vec{W}|=|\vec{V}|.
\end{equation}
%It is easy to derive this result by noticing that, setting
%$\vec{v}_i=\vec{v}_\mathrm{cm}+\vec{u}_i$ and
%$\vec{w}_i=\vec{v}_\mathrm{cm}+\vec{u}'_i$ ($i=1,2$), one has
%$\vec{u}_2=-(m_1/m_2)\vec{u}_1$ (and analogously for $\vec{u}'_i$).
%Substituting in (\ref{kinerg:eq}) one obtains
%\begin{displaymath}
%  \left(m_1+\frac{m_1^2}{m_2}\right)u_1^2=\left(m_1+\frac{m_1^2}{m_2}\right)
%  \left(u_1'\right)^2,
%\end{displaymath}
%from which (\ref{relvec:eq}) follows, since $\vec{V}=-\left(1+m_2/m_1\right) 
%\vec{u}_1$ and $\vec{W}=-\left(1+m_2/m_1\right) \vec{u}'_1$.
Thus the effect of collisions, as seen in the center-of-mass frame, 
amounts to a change in the direction of the relative velocity. It is natural
to assume that the great number of collisions that take part
in the medium make the distribution of $\vec{V}$ isotropic, i.e.,
that the probability that the direction of $\vec{V}$ belongs
to a solid angle $\D\Omega$ depends only on the size
of the solid angle.

We now show that collisions can only reduce the correlation between
$\vec{v}_\mathrm{cm}$ and  $\vec{V}$. 

Let us assume that, at a given time, there is a certain joint
distribution $f(\vec{v}_\mathrm{cm},V,\theta)$ of the
center-of-mass velocity $\vec{v}_\mathrm{cm}$, the
modulus $V$ of the relative velocity and of the
angle $\theta$ between $\vec{V}$ and $\vec{v}_\mathrm{cm}$
for molecule pairs that are about to collide. Thus one has
\begin{displaymath}
  \average{\vec{v}_\mathrm{cm}\cdot\vec{V}}=
  \int \D\vec{v}_\mathrm{cm}\int \D V\int\sin\theta\,\D\theta\,\D\phi
  \;f(\vec{v}_\mathrm{cm},V,\theta)\,v_\mathrm{cm}V\,\cos\theta.
\end{displaymath}
Note that the assumed isotropy of $\vec{V}$ does not imply
that $\average{\cos\theta}=0$, because $\vec{v}_\mathrm{cm}$
determines a special direction. On the other hand, it does imply that
the distribution is invariant with respect to rotations around
$\vec{v}_\mathrm{cm}$, i.e., that the distribution does
not depend on $\phi$.

\begin{figure}[htb]
\psfrag{q}{$\theta_1$}\psfrag{f}{$\phi_1$}
\psfrag{a}{$\alpha$}\psfrag{Q}{$\theta$}
\psfrag{QQ}{$\Theta$}
\begin{center}
\includegraphics[width=8cm]{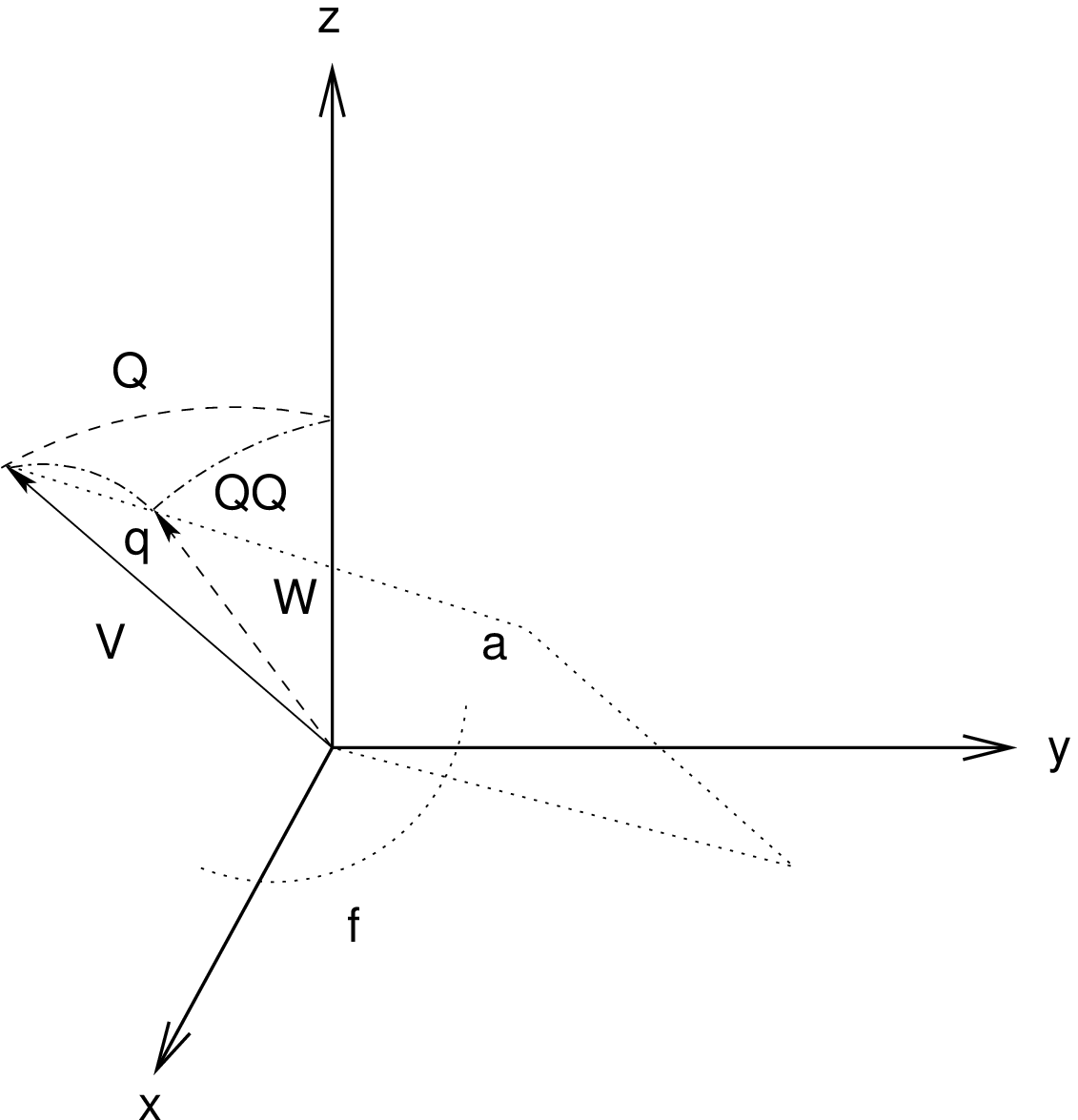}
\end{center}
\caption{Particle scattering. The
$z$-axis lies along $\vec{v}_\mathrm{cm}$ and the
$xz$-plane is defined by
$\vec{v}_\mathrm{cm}$ and $\vec{V}$. We define $\theta$ as the angle
between $\vec{v}_\mathrm{cm}$ and $\vec{V}$. Then $\vec{V}$ and $\vec{W}$
define the plane $\alpha$, which forms the angle $\phi_1$ with the
$xz$-plane. Then, $\theta_1$ is the angle between $\vec{W}$ and
$\vec{V}$. Thus the cosine of the angle $\Theta$
between $\vec{W}$ and $\vec{v}_\mathrm{cm}$ is
given by $\cos\theta \cos \theta_1-
\sin\theta\sin\theta_1\cos\phi_1$.}\label{theta:fig}
\end{figure}

Note that each collision
leaves $\vec{v}_\mathrm{cm}$ and $V$ unchanged, and thus its effect
can be summarized by giving the direction $(\theta_1,\phi_1)$, 
in polar coordinates, of the
relative velocity $\vec{W}$ of the outgoing particles
with respect to the relative velocity $\vec{V}$ 
of the ingoing ones. Invariance
with respect to rotations around $\vec{V}$ intimates
that the probability distribution density $P_V(\theta_1,\phi_1)$ 
of $(\theta_1,\phi_1)$ does not depend on $\phi_1$. 
On the other hand, $P_V(\theta_1,\phi_1)$ is determined only
by the laws of the collision, and should satisfy galelean invariance:
thus it cannot
depend on $\vec{v}_\mathrm{cm}$. We can now evaluate the average
of $\left(\vec{v}_\mathrm{cm}\cdot\vec{V}\right)$ 
by integrating over the relative
direction $(\theta_1,\phi_1)$ of $\vec{W}$ with respect to $\vec{V}$, then
on the relative direction of $\vec{V}$ with respect to $\vec{v}_\mathrm{cm}$,
and finally over $V$ and $\vec{v}_\mathrm{cm}$. Denoting by
$\Theta$ the angle between $\vec{v}_\mathrm{cm}$ and $\vec{W}$, we have
\begin{equation}
  \cos\Theta=\cos\theta \cos\theta_1-\sin\theta\sin\theta_1\cos\phi_1.
\end{equation}
This result can be obtained by considering figure~\ref{theta:fig}. The 
simplest way is to write down the vector $\vec{W}$ as a function of 
$(\theta_1,\phi_1)$, by setting the $z$-axis in the direction
of $\vec{V}$, and then applying a rotation by an angle $\theta$ around
the $y$-axis to the result.
When we average upon $\phi_1$, the second term vanishes. Thus, when 
$\vec{v}_\mathrm{cm}$ and $V$ are fixed, we have
\begin{displaymath}
  \average{\cos\Theta}=\average{\cos\theta\cos\theta_1}=\average{\cos\theta}
  \average{\cos\theta_1},
\end{displaymath}
and thus
\begin{equation}
  \left|\average{\cos\Theta}\right|\le \left|\average{\cos\theta}\right|.
\end{equation}
This holds true also when we average over $\vec{v}_\mathrm{cm}$ and
$V$. Thus
\begin{equation}
  \left|\average{\vec{v}_\mathrm{cm}\cdot\vec W}\right| \le
  \left|\average{\vec{v}_\mathrm{cm}\cdot\vec V}\right|.
\end{equation}

Since collisions reduce correlations, we may conclude that correlations
should vanish at equilibrium:
\begin{equation}
  \average{\vec{v}_\mathrm{cm}\cdot\vec V}_\mathrm{eq}=0.
\end{equation}
From now on, we can follow Feynman's argument. Since
\begin{displaymath}
  \vec{v}_\mathrm{cm}=\frac{m_1\vec{v}_1+m_2\vec{v}_2}{m_1+m_2},
\end{displaymath}
we have
\begin{equation}
  \average{\vec{v}_\mathrm{cm}\cdot\vec{V}}_\mathrm{eq}
  =\frac{\average{{m_2 v_2^2-m_1v_1^2}}_\mathrm{eq}+(m_2-m_1)
    \average{\vec{v}_1\cdot\vec{v}_2}_\mathrm{eq}}{m_1+m_2}.
\end{equation}
The right hand side is proportional to the difference
in kinetic energies of the two kinds of particles if
we assume that the velocities of the colliding particles are
independent, so that $\average{\vec{v}_1\cdot\vec{v}_2}_\mathrm{eq}$ 
vanishes. Now this is one form of the celebrated
\textit{molecular chaos hypothesis}. If it holds at any given
time, we have just shown that the difference between
the mean kinetic energies is reduced. If it holds at
equilibrium, the only possibility is that the difference vanishes:
\begin{equation}
  \frac{1}{2}m_1 \average{v_1^2}_\mathrm{eq}=
  \frac{1}{2}m_2\average{v_2^2}_\mathrm{eq}.
\end{equation}

It is satisfactory to see that the argument does not hold
in one dimension: in this case the only recoil possibility
corresponds to $W=-V$, and thus $|\average{\cos\theta}|=1$.

\section{Discussion}
From a didactical point of view, I guess that the best course would be
to first introduce Maxwell's symmetry arguments leading
to the velocity distribution, and then to demostrate by the present
argument or a similar one that the distribution is left invariant
by collisions, provided that some form of molecular chaos hypothesis
is made. It would be nice to point out that without such an hypothesis
it will always be possible to arrange the molecules in such a way as to
\textit{increase} the kinetic energy difference, e.g., by
reversing the velocities of the particles outgoing a collision.
This should clarify the different roles of mechanical laws
and statistical assumptions in the derivation of the basics
of statistical mechanics.

\ack 
I am grateful to F.~Di~Liberto and G.~Gaeta for their interest in this
work.

\end{document}